# Gap Renormalization in Dirty Anisotropic Superconductors: Implications for the Order Parameter of the Cuprates

R. Fehrenbacher and M.R. Norman

*Science and Technology Center for Superconductivity,*
*Materials Science Division,*
*Argonne National Laboratory,*
*Argonne, Illinois 60439*

**Abstract**. We contrast the effects of non-magnetic impurities on the properties of superconductors having a $d_{x^2-y^2}$ order parameter, and a highly anisotropic s-wave (ASW) gap with the same nodal structure. The non-vanishing, impurity induced, off-diagonal self-energy in the ASW state is shown to gap out the low energy excitations present in the clean system, leading to a qualitatively different impurity response of the single particle density of states compared to the $d_{x^2-y^2}$ state. We discuss how this behaviour can be employed to distinguish one state from the other by an analysis of high-resolution angle-resolved photoemission spectra.

The symmetry of the order parameter (OP) in the high-$T_c$ cuprates is the most crucial and selective single criterion for the determination of their pairing mechanism. A number of recent experiments [1, 2] have provided strong evidence for an anisotropic gap, possibly of $d_{x^2-y^2}$ symmetry. The latter is the most likely candidate for a superconducting OP in purely electronic models, which emphasize the importance of the spin fluctuations induced by the strong Coulomb interactions in the $CuO_2$ planes [3]. However, it has to be kept in mind that most of these experiments probe quantities which involve an average of the modulus of the OP over the Fermi surface. Consequently, the $d_{x^2-y^2}$ gap with its characteristic sign change is practically indistinguishable from a pairing state which has similar modulus, but no sign change, like a strongly anisotropic s-wave (ASW) state, as recently proposed [4]. Experiments with a direct sensitivity to the *phase* of the OP are therefore of greatest importance.

Three phase-sensitive methods for the measurement of the gap have been proposed so far: (a) The magnetic flux dependence of the critical current of a dc SQUID constructed by forming weak links between two s-wave superconductors and the **x** and **y** faces of a single crystal $d_{x^2-y^2}$ superconductor should exhibit a phase shift of $\pi$, *i. e.,* a vanishing critical current for zero flux, opposed to the same configuration using only s-wave superconductors [5, 6]. These experiments have been performed by two independent groups [7] using $YBa_2Cu_3O_{7-\delta}$ as the high-$T_c$-material, both suggesting a $d_{x^2-y^2}$ gap. However, similar interference experiments by other groups lead to results which seem incompatible with the $d_{x^2-y^2}$ symmetry [8]. (b) Measurements of phonon linewidths as a function of temperature were proposed [9] as a possibility to extract the **k** dependence of the OP, but no experimental verification has been reported so far. (c) A sizeable density of midgap surface states was argued to exist in a $d_{x^2-y^2}$ state, whereas they should not be present in any s-wave scenario [10]. Again, there has been no unambigous test for the existence of these states.

In this letter, we propose a third possible phase sensitive method which is capable of distinguishing a $d_{x^2-y^2}$ from an ASW state: it is well known [11] that, opposed to an isotropic s-wave superconductor, scattering by non-magnetic impurities introduces severe pair-breaking in a state with line nodes (such as the $d_{x^2-y^2}$ state), and leads to the appearance of a finite single particle density of states (DOS) at the Fermi energy $\varepsilon_F$, and a rapid suppression of $T_c$.

As we shall show below, this behaviour is in sharp contrast to an ASW state, even if its nodal structure is identical with that of a $d_{x^2-y^2}$ state: due to the additional renormaliza-tion of the off-diagonal self-energy [12] (which vanishes for symmetry reasons in the $d_{x^2-y^2}$ state), isotropic impurity scattering in an ASW superconductor leads to a smearing of the gap anisotropy, and the opening of a gap in the DOS over the whole Fermi surface, even if the gap had line nodes in the pure case. Hence, there is a clear *qualitative* difference in the impurity response of the ASW versus the $d_{x^2-y^2}$ state, which should be detectable in a careful systematic experimental study. We propose angle-resolved photoemission (ARPES) as a suitable probe for this effect. However, care must be taken in the interpretation of the spectra, as we shall discuss at the end of this paper.

We treat the impurity effects in the self-consistent *t*-matrix approximation [13]: The Matsubara components of the single particle propagator in the presence of non-magnetic impurities is expressed in Nambu space as (quantities with a hat represent matrices)

$$\hat{g}(\mathbf{k}, i\omega_n) = \frac{-i\tilde{\omega}_n \hat{\sigma}_0 + \tilde{\Delta}_\mathbf{k} \hat{\sigma}_1 + \tilde{\xi}_\mathbf{k} \hat{\sigma}_3}{\tilde{\omega}_n^2 + \tilde{\Delta}_\mathbf{k}^2 + \tilde{\xi}_\mathbf{k}^2}, \quad (1)$$

with $\tilde{\omega}_n = \omega_n + i\Sigma_0$, $\tilde{\Delta}_\mathbf{k} = \Delta_\mathbf{k} + \Sigma_1$, and $\tilde{\xi}_\mathbf{k} = \xi_\mathbf{k} + \Sigma_3$. Here $\omega_n = \pi(2n+1)T$ is the Matsubara frequency, $T$ the temperature, $\Delta_\mathbf{k}$ the OP (assumed to be real), and $\xi_\mathbf{k}$ the quasiparticle energy measured from $\varepsilon_F$. The self-energy $\hat{\Sigma} = \sum_{j=0}^{3} \Sigma_j \hat{\sigma}_j$ (and all other matrix quantities) is expanded in terms of the unit and Pauli matrices $\hat{\sigma}_0 \ldots \hat{\sigma}_3$ and is given by $\hat{\Sigma}(i\omega_n) = \Gamma \hat{T}(i\omega_n)$, where $\hat{T}$ is the *t*-matrix, $\Gamma = n_i/(\pi N_0)$, $n_i$ the impurity concentration, and $N_0$ the DOS at $\varepsilon_F$ in the normal state. Finally the *t*-matrix is self-consistently determined from

$$\hat{T}(i\omega_n) = \hat{K}_n + \hat{K}_n \hat{G}(i\omega_n) \hat{T}(i\omega_n), \quad (2)$$

where $\hat{G}(i\omega_n) = (\pi N_0)^{-1} \sum_\mathbf{k} \hat{g}(\mathbf{k}, i\omega_n)$, and the normal state K-matrix is parametrized by an *s*-wave scattering phase shift $\delta_0$, $\hat{K}_n = -\hat{\sigma}_3/c$, $c = \cot \delta_0$, which measures the scattering strength ($c = 0$ unitary limit, $c \gg 1$ Born limit). Assuming particle-hole symmetry, it can be shown [13] that $G_2 = G_3 = 0$. Then the non-vanishing *t*-matrix components are given by

$$T_0 = \frac{G_0}{c^2 - G_0^2 + G_1^2} \qquad T_1 = \frac{-G_1}{c^2 - G_0^2 + G_1^2}. \quad (3)$$

A self-consistent solution to Eqs.(1-3) is obtained if $\Delta_\mathbf{k}$ satisfies the gap equation

$$\Delta_\mathbf{k} = -T \sum_{\omega_n} \sum_{\mathbf{k}'} V_{\mathbf{k}\mathbf{k}'} \frac{\tilde{\Delta}_{\mathbf{k}'}}{\tilde{\omega}_n^2 + \tilde{\Delta}_{\mathbf{k}'}^2 + \tilde{\xi}_{\mathbf{k}'}^2}, \quad (4)$$

where $V_{\mathbf{kk'}}$ is the pair potential.

In the following, we consider a two-dimensional electron system with a circular Fermi surface, and a separable pair potential $V_{\mathbf{kk'}} = V\eta(\phi)\eta(\phi')$ ($\phi$ the polar angle in the plane). We consider two possibilities for the angular function $\eta(\phi)$: (i) The dependence $\eta_d(\phi) = \cos 2\phi$, which leads to the characteristic $d_{x^2-y^2}$ gap, $\Delta(\phi) = \Delta\cos 2\phi$, and (ii) $\eta_s(\phi) = |1 - \frac{4}{\pi}\phi|$, $0 \leq \phi \leq \pi/2$, periodically continued to the interval $[\pi/2, 2\pi]$, i. e., a sawtooth function with period $\pi/2$. The latter case leads to a particular ASW state with $\Delta(\phi) = \Delta\eta_s(\phi)$. This is the simplest form of an ASW state with the same nodal structure as the $d_{x^2-y^2}$ OP, and would therefore be practically indistinguishable from it in all experiments which do not probe the phase of the OP explicitly. In reality, it is very unlikely that an ASW OP would have real line nodes, since there is no symmetry reason for it to vanish. However, the existence of a small isotropic part in the gap would not alter the conclusions of this paper. The form of $\eta_s(\phi)$ was chosen to emphasize the different effects of impurities in the two cases, and for its mathematical simplicity, which allows for the solution of Eqs. (1-4), at all temperatures and impurity concentrations.

The fundamental difference between the two states appears in the presence of non-magnetic impurities, signaled by the $\hat{\sigma}_1$-component of the integrated single particle propagator

$$G_1(i\omega_n) = \frac{1}{\pi N_0} \sum_{\mathbf{k}} \frac{\widetilde{\Delta}_{\mathbf{k}}}{\widetilde{\omega}_n^2 + \widetilde{\Delta}_{\mathbf{k}}^2 + \widetilde{\xi}_{\mathbf{k}}^2}. \tag{5}$$

The symmetry of the $d_{x^2-y^2}$ gap makes the sum over $\mathbf{k}$ vanish, and therefore $G_1 = \Sigma_1 = 0$, i. e., $\widetilde{\Delta}_{\mathbf{k}} = \Delta_{\mathbf{k}}$, whereas in the ASW state, $G_1$ and $\Sigma_1$ are finite. This leads to an *isotropic* (due to the assumption of pure $s$-wave scattering) renormalization of the off-diagonal self-energy by the impurities, $\widetilde{\Delta}_{\mathbf{k}} = \Delta_{\mathbf{k}}+\Sigma_1$. Below, we shall discuss how this difference in the gap renormalization affects several observable quantities, and how this could be used to distinguish the two gap symmetries.

We have self-consistently solved Eqs. (1-4) at all temperatures for both OP symmetries, and compared two values for the scattering strength $c = 0, 1$, representative of strong and intermediate scattering. First consider the $T_c$ suppression: In both cases, we obtain the standard Abrikosov-Gorkov result [14], $\ln(T_{c0}/T_c) = a[\Psi(1/2+\alpha/(2\pi T_c))-\Psi(1/2)]$, where $\alpha = \Gamma/(c^2+1)$ is the pair-breaking parameter. In the $d_{x^2-y^2}$ case, $a = 1$, and we obtain the critical concentration $n_c = \pi^2(c^2 + 1)N_0 T_{c0}/(2e^\gamma)$ at which $T_c$ vanishes, whereas in the ASW case, $a = 1/4$, and $n_c = \infty$. In Fig. 1(a), we show the dependence of $T_c$ on $\Gamma$ (all energies are scaled in terms of the maximum gap value $\Delta_{00} = \Delta(\phi = 0, T = 0, n_i = 0)$). The value of $N_0$ in each $CuO_2$ plane can be estimated to be of the order of 1 - 2 eV$^{-1}$ per Cu, which leads to the relation $\Gamma = vn_i$, where $v \approx 2$ meV, and $n_i$ is in percent. The initial decrease in $T_c$ is given by $T_{c0} - T_c = a\alpha\pi/4$, i.e. the suppression in a $d_{x^2-y^2}$ superconductor is four times faster than for our particular choice of ASW state. But note that in the ASW scenario, the coefficient $a$ is non-generic.

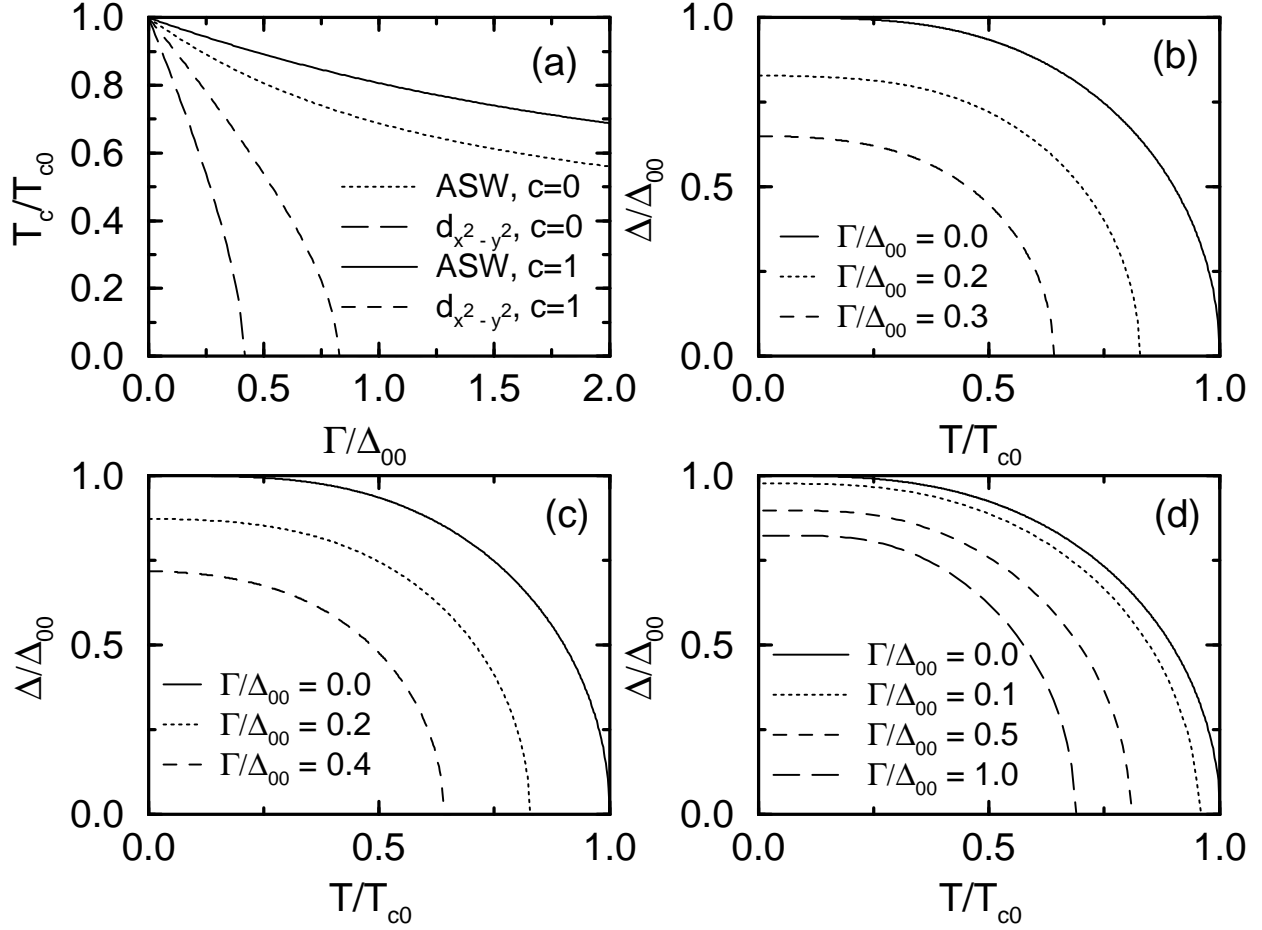

**Figure** 1: (a) The $T_c$-suppression as a function of impurity concentration $n_i$ for the $d_{x^2-y^2}$ and the ASW state at two scattering strengths $c = 0, 1$. The temperature dependence of the $d_{x^2-y^2}$ OP for different $n_i$, at $c = 0$ (b), and $c = 1$ (c). (d) The temperature dependence of the ASW OP for different $n_i$, at $c = 0$.

Figs. 1(b-d) show the temperature dependence of the OP for different impurity concentrations. In the $d_{x^2-y^2}$ case, Figs. 1(b,c), the suppression of the zero temperature gap is very similar to the $T_c$ suppression, whereas in the ASW case, Fig. 1(d), it is suppressed slower than $T_c$.

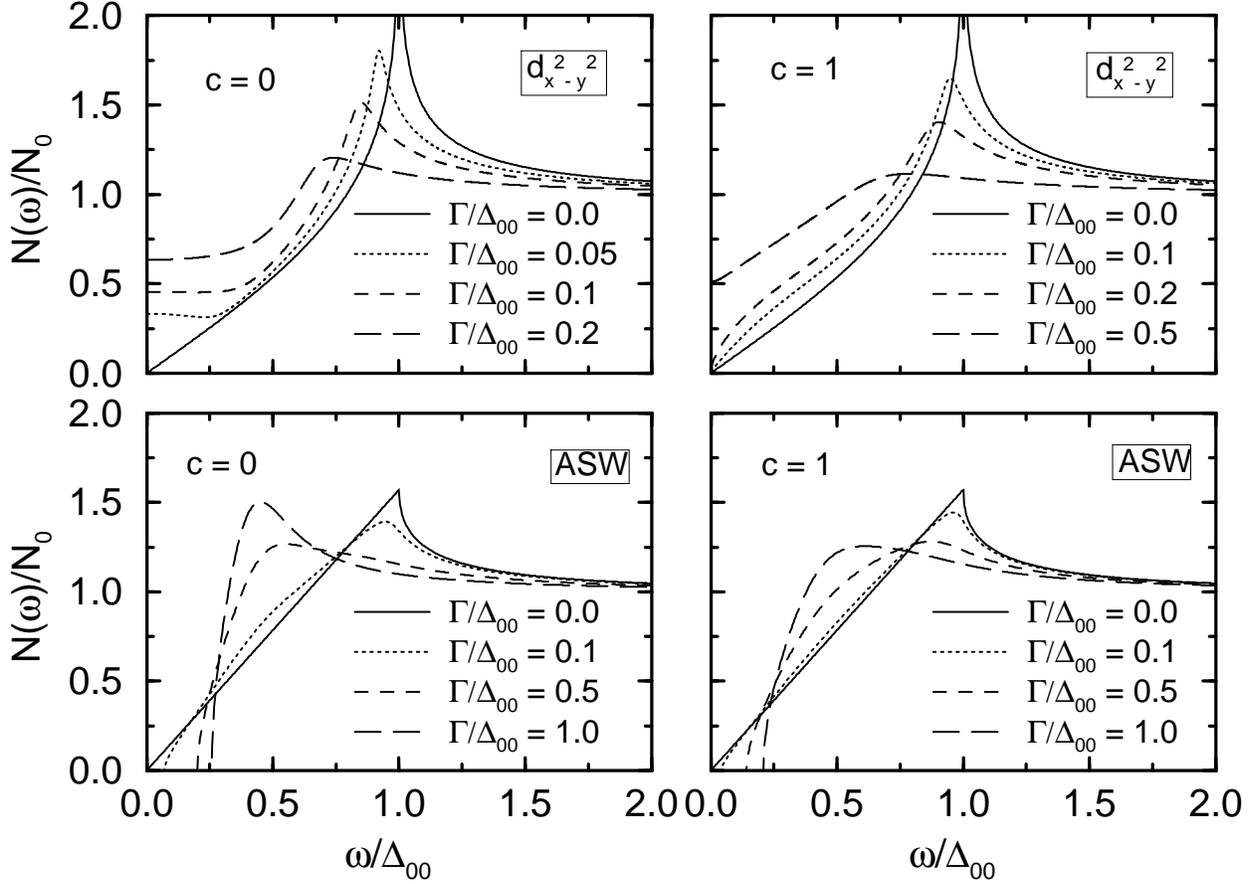

**Figure** 2: A comparison of the DOS at $T = 0$ for the $d_{x^2-y^2}$ OP and the ASW OP for two scattering strengths ($c = 0, 1$) at different values of the impurity concentration.

Next we consider the DOS, which is given by $N(\omega)/N_0 = -\Im m G_0(i\omega_n)|_{i\omega_n=\omega+i\delta}$. Fig. 2 shows the impurity dependence of the DOS for the $d_{x^2-y^2}$ and the ASW state for the two scattering strengths $c = 0, 1$ at $T = 0$. Note that the frequency in these plots is scaled in terms of $\Delta_{00}$, *i. e.*, the gap suppression is taken into account. In the pure case, $N(\omega)$ is linear at small frequencies for both states. The logarithmic singularity at $\omega = \Delta_{00}$ in the $d_{x^2-y^2}$ case is replaced by a cusp in the ASW state, due to our choice of $\eta_s(\phi)$. As soon as impurities are doped into the system, the difference between the two states becomes apparent: Whereas, in the $d_{x^2-y^2}$ state, the impurities lead to excitations at zero energy over the *whole* Fermi surface (as verified by a calculation of the angle resolved DOS, $N(\phi, \omega)/N_0 \equiv \Im m\{i\widetilde{\omega}_n[4\pi^2(\widetilde{\omega}_n^2+\widetilde{\Delta}_\phi^2)]^{-1/2}\}|_{i\omega_n=\omega+i\delta}$), their effect in the ASW state is to induce a finite *energy gap*, which initially grows substantially as a function of the impurity concentration $n_i$. For very large $n_i$, the gap anisotropy in the ASW state becomes completely smeared out by the isotropic scattering, and the DOS approaches the familiar shape of an isotropic s-wave gap. This should also have the consequence of activated behaviour in thermodynamic and transport properties of dirty samples at low temperatures, in contrast to the power laws predicted for clean [15] or dirty [16] $d_{x^2-y^2}$ states. Note also that the resonance behaviour found in the unitary scattering limit ($c = 0$) of the $d_{x^2-y^2}$ state at low-frequency is cut off in the ASW state. This is due to the fact that in the latter case, the finite gap renormalization prevents the denominator of $T_0$ in Eq. 3 to become singular. Consequently there is no qualitative difference between the results obtained in the Born ($c \gg 1$) and unitary scattering limit, and all the results can essentially be parametrized by the effective normal state inverse scattering rate $\Gamma/(c^2 + 1)$.

The indication of an anisotropic gap found by ARPES [2] motivated us to calculate the impurity dependence of the spectral function $A(\mathbf{k}, \omega)$, which is defined as

$$A(\mathbf{k}, \omega) = -\frac{\text{sgn}\omega}{\pi} \Im \left\{ \widehat{g}_{11}(\mathbf{k}, i\omega_n)|_{i\omega_n=\omega+i\delta} \right\}, \tag{6}$$

where $\widehat{g}_{11}$ is the 11 component of the single particle matrix propagator. In the pure case of the BCS models we consider, the spectral function consists of the familiar two $\delta$ functions at the quasiparticle energies $\omega = \pm E_\mathbf{k}$ [17]

$$A(\mathbf{k}, \omega) = u_\mathbf{k}^2 \delta(\omega - E_\mathbf{k}) + v_\mathbf{k}^2 \delta(\omega + E_\mathbf{k}), \tag{7}$$

where $E_\mathbf{k} = \sqrt{\xi_\mathbf{k}^2 + \Delta_\mathbf{k}^2}$, $u_\mathbf{k}^2 = (1 + \xi_\mathbf{k}/E_\mathbf{k})/2$, $v_\mathbf{k}^2 = (1 - \xi_\mathbf{k}/E_\mathbf{k})/2$. Hence the quasiparticle peaks shift by an energy of $E_\mathbf{k} - |\xi_\mathbf{k}|$ when going from the normal to the superconducting state.

Before we proceed with the results obtained in the presence of impurities, let us briefly discuss the relation of the spectral function to the intensity $I(\mathbf{k}, \omega)$ measured in an ARPES experiment. For a perfectly flat surface, the photoexcited electron preserves its momentum parallel to the surface, when leaving the sample. In the case of the cuprate superconductors, one therefore cleaves the crystal with a surface parallel to the **a** − **b** plane, so that momentum information is obtained about the relevant $CuO_2$ planes. Then the intensity becomes a function of the normal quasiparticle energy $\xi$, the angle $\phi$, and $\omega$. The energy difference of the outgoing electron to the incoming photon gives information about the negative frequency part of $A(\phi, \xi, \omega)$. The substantial background intensity which is usually observed is caused by multiple inelastic scattering of some photoexcited electrons before they leave the sample.

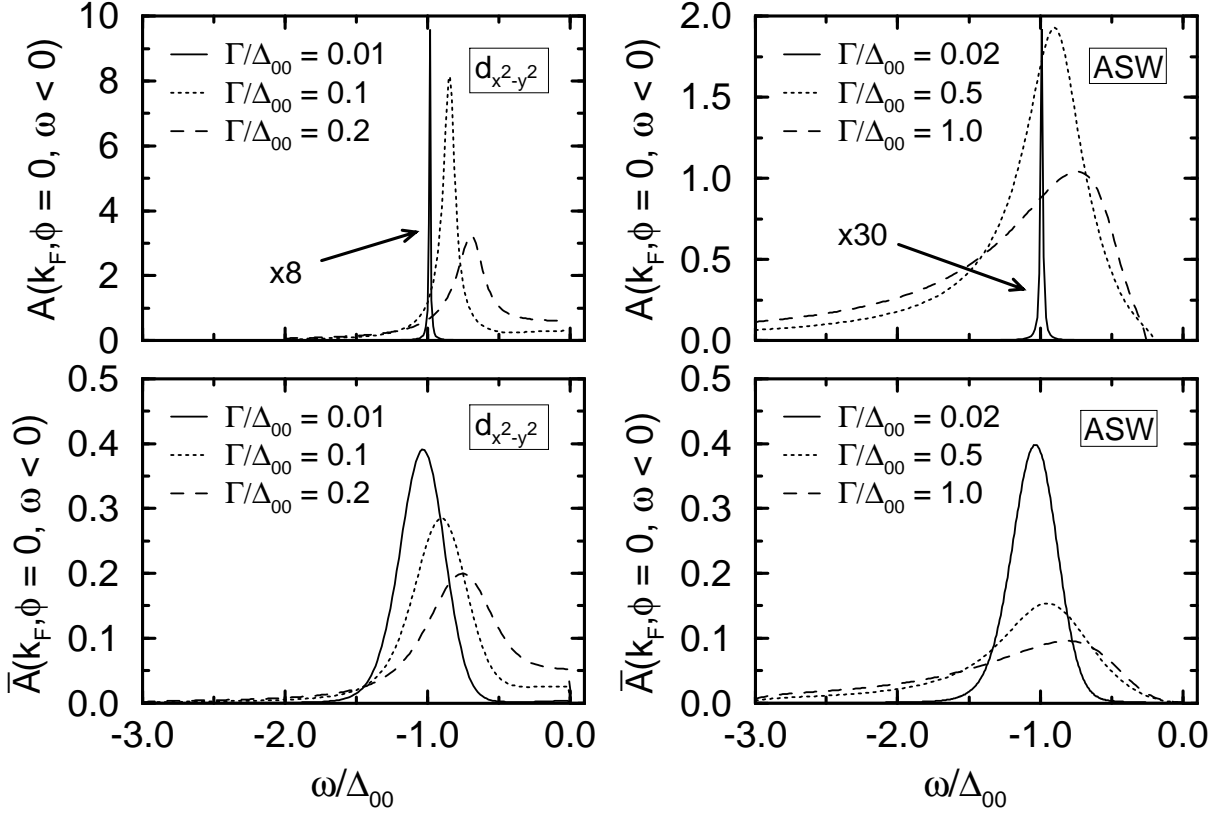

**Figure** 3: The averaged (bottom curves) and true (top curves) spectral functions for the $d_{x^2-y^2}$ and the ASW OP at $\xi = 0$, $\phi = 0$ for varying impurity concentration in the unitary limit.

For a comparison of theoretical results to the measured spectra, it is important to take into account the finite experimental resolution, both in energy and momentum. We have simulated the effect of finite angular resolution by a Gaussian width $\Delta_\xi$ for the normal quasiparticle energy, and similarly the finite energy resolution by a corresponding width $\Delta_\omega$ for the frequency. We then argue that the measured intensity should be proportional to $A(\phi, \xi, \omega < 0)$ integrated with these probability distributions.

In Fig. 3, we plot the spectra for $A(k_F, \phi = 0, \omega < 0)$, *i. e.*, along the direction of the gap maximum, calculated as a function of impurity concentration in the unitary limit for both states. We always plot the spectra for $\xi = 0$, *i. e.*, at the Fermi surface, and $T = 0$. The top and bottom curves are obtained before and after the averaging process respectively. The resolution widths were chosen as $\Delta_\omega/\Delta_{00} = 0.2$, and $\Delta_\xi/\Delta_{00} = 0.5$. Assuming a value of $\Delta_{00} \approx 30$meV, these values correspond to the best available resolutions on current spectrometers.

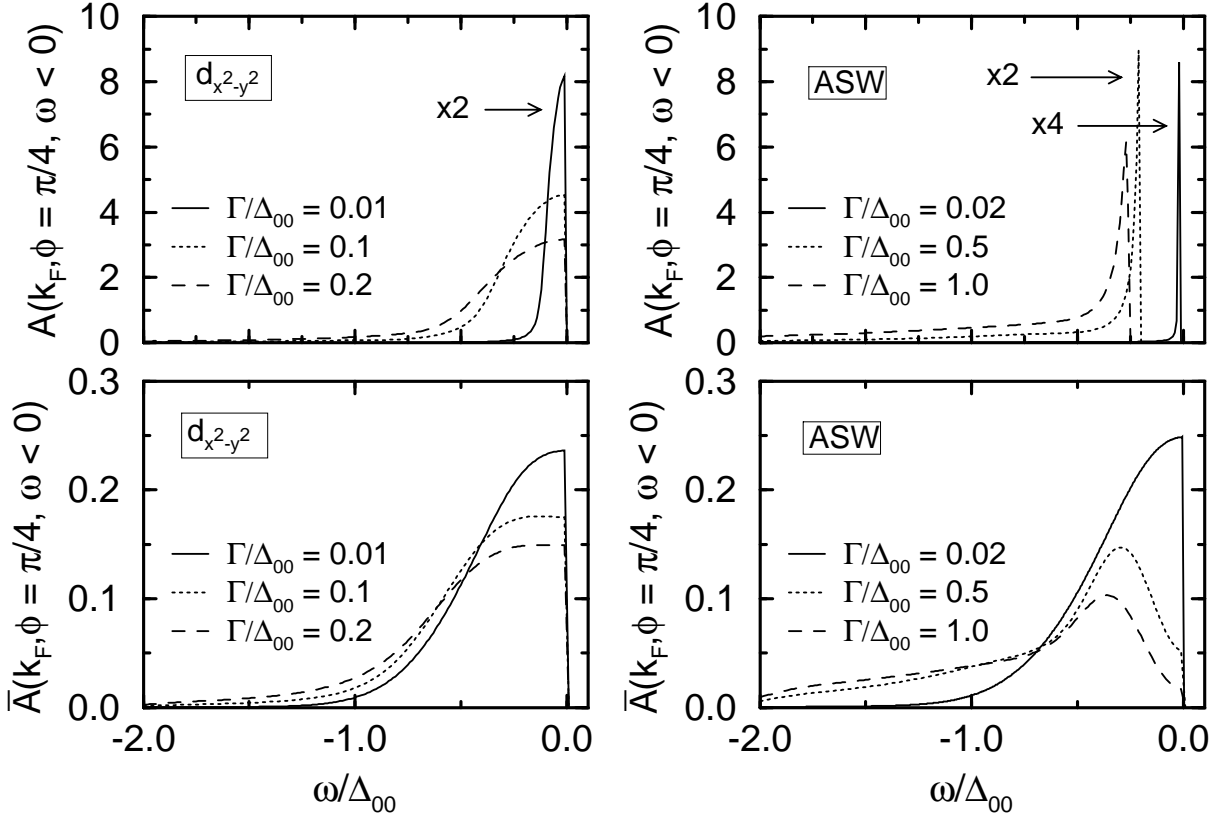

**Figure** 4: The averaged (bottom curves) and true (top curves) spectral functions for the $d_{x^2-y^2}$ and the ASW OP at $\xi = 0$, $\phi = \pi/4$ for varying impurity concentration in the unitary limit.

For increasing impurity concentration the effect on the spectra is qualitatively similar for both states: the quasiparticle peaks shift to smaller energy (due to the reduction in $\Delta_0$), and they become substantially broadened. Around zero frequency, the spectra are different: Similar to the total DOS, the spectral weight is non-zero in the $d_{x^2-y^2}$ case, whereas there is a true gap in the ASW state around $\omega = 0$. However the broadened spectra indicate that this difference would be difficult to detect, especially since it occurs in the vicinity of $\varepsilon_F$, where the ARPES experiments suffer from a sharp drop in intensity.

Fig. 4 shows the spectra along the node directions calculated with the same parameters as in Fig. 3. Along this direction, the two states show a clear qualitative difference: for the $d_{x^2-y^2}$ state, the only effect of an enhanced impurity concentration is the broadening of the peak which remains right at $\varepsilon_F$. In the ASW case, however, the opening up of a gap in the presence of impurities makes the peak shift away from $\varepsilon_F$. When looking at the averaged spectra, we notice that the shift of the peak is still clearly visible, even though its sharp cutoff is smoothed out by the finite resolution. This shows that a clear distinction of the ASW from the $d_{x^2-y^2}$ state should be detectable in a systematic analysis of high-resolution ARPES spectra taken at the possible line nodes as a function of non-magnetic impurity concentration. Note that a similar effect would happen if the OP in the ASW state had a small isotropic part.

Since the experiments are done at finite temperature, it is important to divide the spectra by the Fermi occupation factor (after a careful background subtraction) for a comparison of the data to our theoretical curves. Figs. 1(b-d) show that the reduction of $\Delta$ by increasing the temperature is negligible if $T/T_c < 0.2$ (as in actual measurements), so that our $T = 0$ results should be applicable when thermal occupation of quasiparticle states with $E_\mathbf{k} > 0$ is taken into account.

A final comment concerns the data published by Shen *et al.* [2]. They showed that the anisotropy of their extracted gap values decreases as a function of time after the cleavage of the sample. If this aging would be due to the creation of non-magnetic impurities, *e. g.*, by the diffusion of oxygen, then the effect could be understood in terms of an ASW state as we showed above. It is also possible that it is due to a degradation of the surface quality, as Shen *et al.* argue, in which case the resulting poorer **k** resolution could be responsible for the effect. But in this case, one might expect (i) an enhanced width of the superconducting quasiparticle peak which was not observed, and (ii) that the poorer **k**-resolution should also wash out the **k**-dispersion in the normal state data. A sytematic study of bulk irradiated samples would certainly be more conclusive, and is therefore highly desirable. However it has to be made sure that the radiation damage does not lead to localized magnetic moments, or a shift in the chemical potential, since this would mask the effects we were discussing.

In conclusion, we have compared the effect of non-magnetic impurities on the properties of superconductors with a $d_{x^2-y^2}$ versus a highly anisotropic s-wave OP, both with the same nodal structure. We have shown that even though the two states are indistinguishable in experiments which probe Fermi surface averages of the OP, their behaviour is qualitatively different in the presence of non-magnetic impurities. The difference comes from the non-vanishing, impurity induced, off-diagonal self-energy in the ASW state, which leads to the opening up of a true gap in the single particle excitations of dirty samples. In particular, we discussed how this effect could be observed in ARPES experiments.

We acknowledge useful discussions with J.C. Campuzano, C. Sá de Melo, R. J. Radtke and Z.-X. Shen, as well as the financial support of the National Science Foundation through grant NSF-DRM-91-20000 (R.F.), and of the U.S. Department of Energy, Office of Basic Energy Sciences, under Contract No. W-31-109-ENG-38 (M.R.N.).

# References

[1] For a summary, see B. Goss Levi, Physics Today **46** No. 5, 17 (1993).

[2] Z.-X. Shen *et al.*, Phys. Rev. Lett. **70**, 1553 (1993).

[3] D. J. Scalapino, J. E. Loh, and J. E. Hirsch, Phys. Rev. B **34**, 8190 (1986).

[4] S. Chakravarty *et al.*, Science **261**, 337 (1993).

[5] V. B. Geshkenbein and A. I. Larkin, Pis'ma Zh. Eksp. Teor. Fiz. **43**, 306 (1986) [Sov. Phys. JETP Lett. **43**, 395 (1986)]; V. B. Geshkenbein, A. I. Larkin and A. Barone, Phys. Rev. B **36**, 235 (1987).

[6] M. Sigrist and T. M. Rice, J. Phys. Soc. Jpn. **61**, 4283 (1992).

[7] D. A. Wollman *et al.*, Phys. Rev. Lett. **71**, 2134 (1993); D. A. Brawner and H. R. Ott, preprint (1994).

[8] P. Chaudhari and S.-Y. Lin, Phys. Rev. Lett. **72**, 1084 (1994); A. G. Sun *et al.*, preprint (1993).

[9] M. E. Flatté, S. Quinlan, and D. J. Scalapino, Phys. Rev. B **48**, 10626 (1993).

[10] C.-R. Hu, Phys. Rev. Lett. **72**, 1526 (1994).

[11] L. P. Gorkov, Pis'ma Zh. Eksp. Teor. Fiz. **40**, 351 (1984) [Sov. Phys. JETP Lett. **40**, 1155 (1985)]; K. Ueda and T. M. Rice, in *Theory of Heavy Fermions and Valence Fluctuations*, edited by T. Kasuya and T. Saso (Springer, Berlin, 1985), p. 267.

[12] T. Tsuneto, Prog. Theor. Phys. **28**, 857 (1962); D. Markowitz and L. P. Kadanoff, Phys. Rev. **131**, 563 (1963); P. Hohenberg, Zh. Eksp. Teor. Fiz. **45**, 1208 (1963) [Sov. Phys. JETP **18**, 834 (1964)].

[13] P. J. Hirschfeld, P. Wölfle, and D. Einzel, Phys. Rev. B **37**, 83 (1988).

[14] A. Abrikosov and L. P. Gorkov, Zh. Eksp. Teor. Fiz. **39**, 1781 (1960) [Sov. Phys. JETP **10**, 593 (1960)].

[15] F. Gross *et al.*, Z. Phys. B **64**, 175 (1986).

[16] P. J. Hirschfeld and N. Goldenfeld, Phys. Rev. B **48**, 4219 (1993).

[17] J. R. Schrieffer, *Theory of Superconductivity*, Benjamin Cummings, Reading, Massachusetts, 1964.